\documentclass[aps,prl,reprint]{revtex4-1}
\usepackage{amsmath,amssymb,graphicx}

\begin{document}

\title{Asymptotic Equation for Zeros of Hermite Polynomials from the Holstein-Primakoff Representation}
\author{Lucas Kocia}
\affiliation{Department of Chemistry and Chemical Biology, Harvard University, Cambridge, Massachusetts 02138}
\email[To whom correspondence should be addressed: E-mail:]{lkocia@fas.harvard.edu.}
\begin{abstract}
The Holstein-Primakoff representation for spin systems is used to derive expressions with solutions that are conjectured to be the zeros of Hermite polynomials \(H_n(x)\) as \(n \rightarrow \infty\). This establishes a correspondence between the zeros of the Hermite polynomials and the boundaries of the position basis of finite-dimensional Hilbert spaces.
\end{abstract}
\maketitle

The Hermite polynomials are prevalent in many fields. They can be defined as
\begin{equation}
  H_n(x) = (-1)^n e^{x^n} \frac{d^n}{dx^n} \left( e^{-x^2} \right).
\end{equation}
In the physics community, they are perhaps best recognized as the Gaussian-weighted eigenfunctions (in position representation) of the quantum harmonic oscillator (with \(\hbar = m = \omega = 1\), a convention that will be used for the rest of the paper):
\begin{equation}
  \frac{1}{2} \left( x^2 - \frac{d^2}{d x^2} \right) e^{-\frac{x^2}{2}} H_n(x) = \left( n+\frac{1}{2} \right) e^{-\frac{x^2}{2}} H_n(x),
\end{equation}
As such, they are orthogonal over the Gaussian-weighted whole domain, \(\int^\infty_{-\infty} H_m(x) H_n(x) e^{-x^2} dx = \sqrt{\pi}2^n n! \delta_{n m}\). This last property allows their use in Gaussian quadrature, a useful and popular numerical integration technique where \(\int^\infty_{-\infty} f(x) dx\) is approximated as \(\sum^n_{j=1} e^{-x_j^2} f(x_j)\) where \(x_j\) are the zeros of \(H_n(x)\) and \(f(x)\) is a well-behaved function. For this and many other reasons, an analytic formula for the asymptotic zeros of Hermite and other orthogonal polynomials has been a subject of much interest\cite{Nevai79, Ullman80, Mhaskar84, Gonchar86, Gawronski87, Dominici07, Elbert08}, especially in the applied mathematics community and the field of approximation theory.

In this paper, I examine the position state representation of the eigenstates of finite dimensional \(S\)-spin systems, as expressed in the Holstein-Primakoff transformation. As \(S \rightarrow \infty\), the system becomes the infinite dimensional harmonic oscillator. This association allows me to derive the simple main results presented in eqs~\ref{eq:evenzeros} and \ref{eq:oddzeros}, with solutions that I conjecture become the asymptotic zeros of the Hermite polynomials (as \(n \rightarrow \infty\)). Furthermore, I numerically show that this convergence is rather quick and so the expressions can frequently be used, in many instances of finite-precision application, as the effective zeros of \(H_n(x)\) with finite \(n\), such as in applications of Gaussian quadrature. In a more aesthetic sense, these results also establish a beautiful correspondence between the boundaries of equal area partitions of circles with radii that are increasing in a certain manner and the Hermite polynomial zeros.

\vspace{5pt}
Spin systems are defined by the fundamental commutation relations between operators \(\hat S^z\), \(\hat S^+\) and \(\hat S^-\) :

\begin{equation}
  \left[ \hat S^z, \hat S^+ \right] = \hat S^+, \hskip 5pt \left[ \hat S^z, \hat S^- \right] = \hat S^-, \hskip 5pt \left[ \hat S^+, \hat S^- \right] = 2 \hat S^z.
\end{equation}

Associating a spin with a boson \(c^\dagger\), Holstein and Primakoff showed that to satisfy these commutation relations, the operators can be expressed as\cite{HolPrim40}

\begin{equation}
  \hat S^z = \hat c^\dagger \hat c - S,
\end{equation}
\begin{equation}
  \hat S^+ = \hat c^\dagger \sqrt{2 S - \hat c^\dagger \hat c}, \hskip 5pt \mbox{and} \hskip 5pt \hat S^- = \sqrt{2 S - \hat c^\dagger \hat c} \, \hat c.
  \label{eq:HKladder}
\end{equation}

This is a very useful association and has found many applications in the condensed matter field's study of many-body spin systems.  Each boson excitation represents the ``ladder up'' finitesimal excitation away from the spin's extremal \(S\) state. The Hilbert space is finite-dimensional and possesses \(2S+1\) states \(\{-S,-S+1, \ldots, S\}\). In fact, considering eq.~\ref{eq:HKladder} it is clear that the Hilbert space outside this defined space is not even Hermitian. 

\begin{figure}[h]
\includegraphics[scale=0.3]{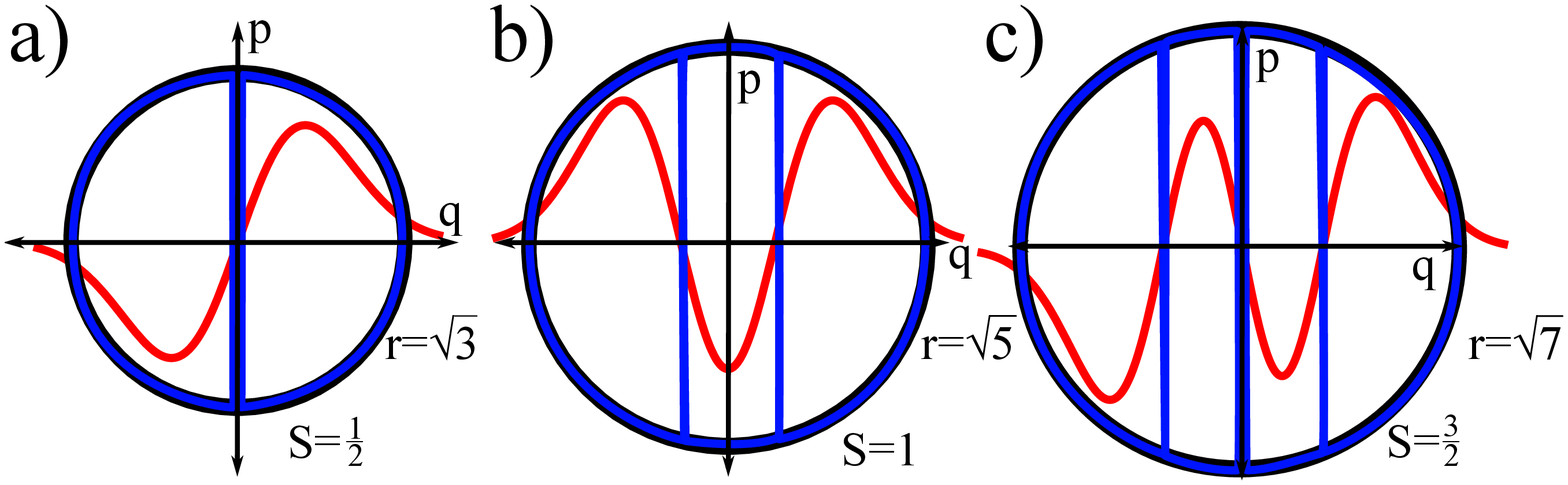}
\caption{The \(q\)-basis representation of a) \(S = \frac{1}{2}\), b) \(S = 1\) and c) \(S = \frac{3}{2}\) systems is shown. The radius of the Hilbert space's domain is equal to \(\sqrt{4S+1}\) and so grows along with the number of allowed basis elements.}
\label{fig:qbasis}
\end{figure}

Transforming from the Holstein-Primakoff bosonic representation to position (and its conjugate momentum) space (using the relations \(c^\dagger = \frac{1}{\sqrt{2}} \left( \hat q - i \hat p \right)\) and \(c = \frac{1}{\sqrt{2}} \left( \hat q + i \hat p \right)\)) reveals that the trivial Hamiltonian is the harmonic oscillator: \(\hat H = \hat S_z = \frac{1}{2} \left( \hat q^2 + \hat p^2 \right) - \left(S+\frac{1}{2}\right)\). Moreover, transformation of the \(\hat S^+\) and \(\hat S^-\) in eq.~\ref{eq:HKladder} reveals that the Hilbert space spans the domain \(r^2 \equiv p^2 + q^2 \le \sqrt{4 S + 1}\). Just as in the \(S_z\) representation, \(2 S\) states all with the same area must exist within this domain. Fig.~\ref{fig:qbasis} sketches out what they look like for the \(\{S=\frac{1}{2}, S = 1, S = \frac{3}{2}\} \)-spin systems.

For a particular \(S\)-spin system, the lowest eigenstate must have the same sign at all \(q\)-basis elements since it must be nodeless. On the other hand, the highest eigenstate must have \(n-1\) nodes and so the \(q\)-basis elements must alternate in sign such that the eigenfunction passes through zero between them. This latter behavior is sketched in fig.~\ref{fig:qbasis} in red by the Hermite polynomial \(H_n(x)\) denoting the value of the overlying \(q\)-basis element for the highest eigenstate.

For \(S \rightarrow \infty\), the Hilbert space becomes infinite-dimensional and the Hamiltonian becomes that of the harmonic oscillator defined over \((p,q) \in \mathbb{R}^2\) with the associated eigenfunctions proportional to \(e^{-\frac{x^2}{2}} H_n(x)\). It therefore follows that as \(S \rightarrow \infty\), the boundaries of the \(q\)-basis elements become the zeros of the Hermite polynomial \(H_n(x)\) where \(n = 2S\) since the highest eigenstate must still have alternating sign with each \(q\)-basis element.

\begin{figure}[h]
\includegraphics[scale=0.4]{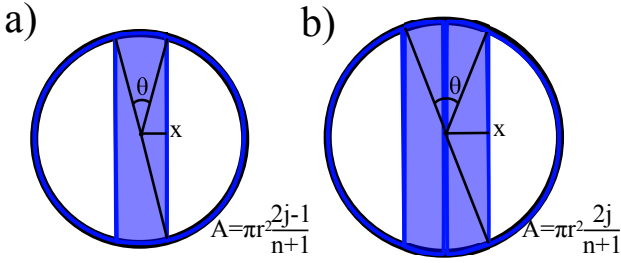}
\caption{The area of the central a) \(2j-1\) or b) \(2j\) \(q\)-basis elements that approximately determine the \(j\)th zero of the Hermite polynomial \(H_n(x)\) for \(n\) even and odd respectively is shaded in blue. The approximate \(j\)th zero is at the right boundary of these regions.}
\label{fig:areas}
\end{figure}

Hermite polynomial zeros \(x_j\) are real and symmetric around \(x = 0\). To determine these boundary points, the \(2S\)-dimensional Hilbert space's circular shape in position space can be exploited. For even \(2S\), the area of the all the \(q\)-basis elements up until the \(j\)th boundary (measuring from the origin) is \(\pi r^2 \frac{2j-1}{n+1}\). For odd \(2S\), the area is \(\pi r^2 \frac{2j}{n+1}\). This is illustrated in fig.~\ref{fig:areas}.

Using simple relations for the area of circle sectors and rectangles, it is possible to relate these \(q\)-basis element areas to \(x_j\); The equation involving the approximate zeros of Hermite  polynomials \(H_n\) with \(n\) even is:
{
\begin{eqnarray}
\frac{(2 j-1) \pi }{n+1} &=& \sin \theta + \theta,
\label{eq:evenzeros}
\end{eqnarray}
}%

while for odd \(n\) it is:
{
\begin{eqnarray}
\frac{2 j \pi }{n+1} &=& \sin \theta + \theta,
\label{eq:oddzeros}
\end{eqnarray}
}%

where \(\theta = 2 \sin^{-1} \frac{x_j}{r}\) and \(r = \sqrt{2n + 1}\).

\begin{figure}[h]
\includegraphics[scale=0.5]{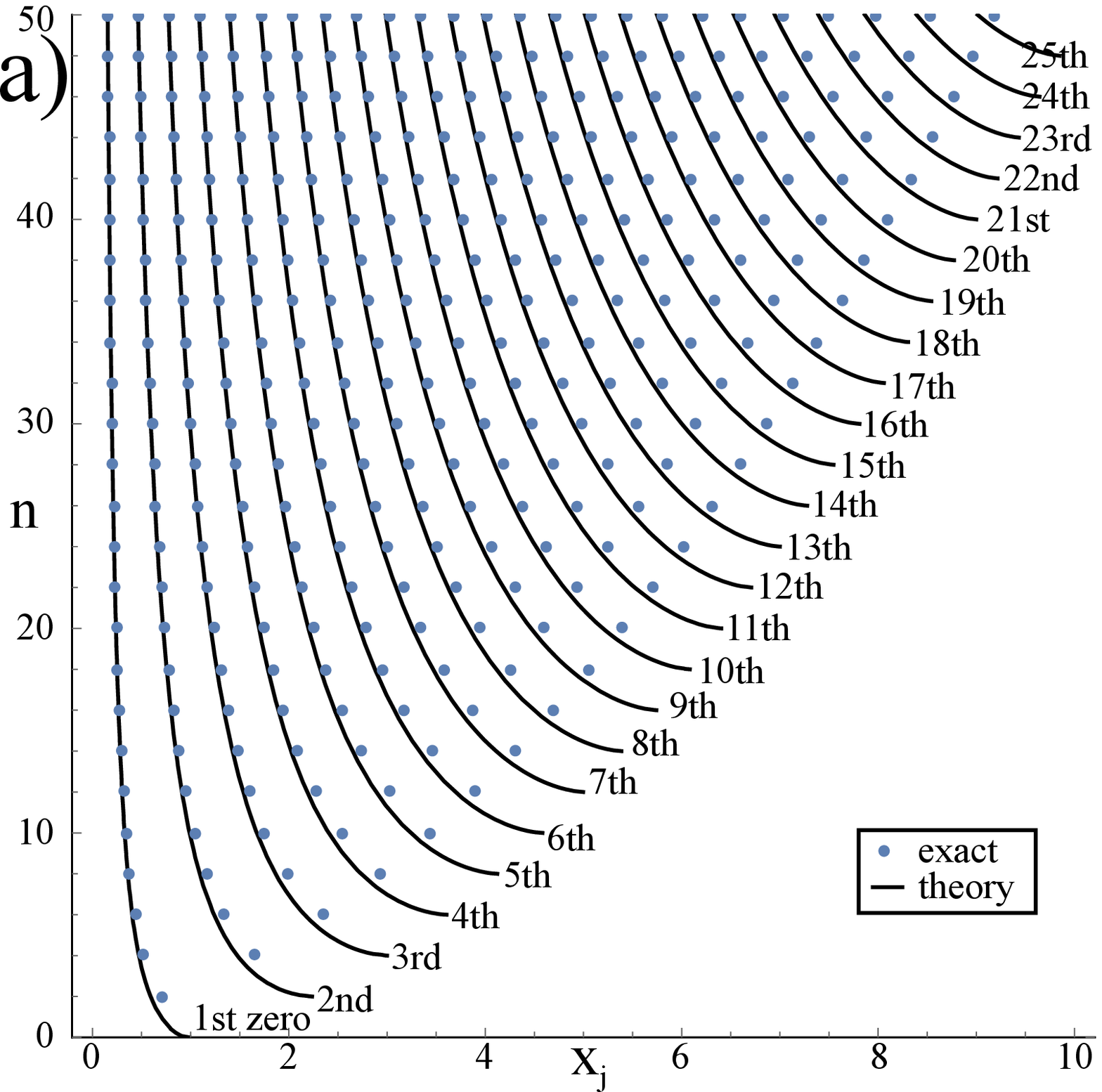}
\includegraphics[scale=0.5]{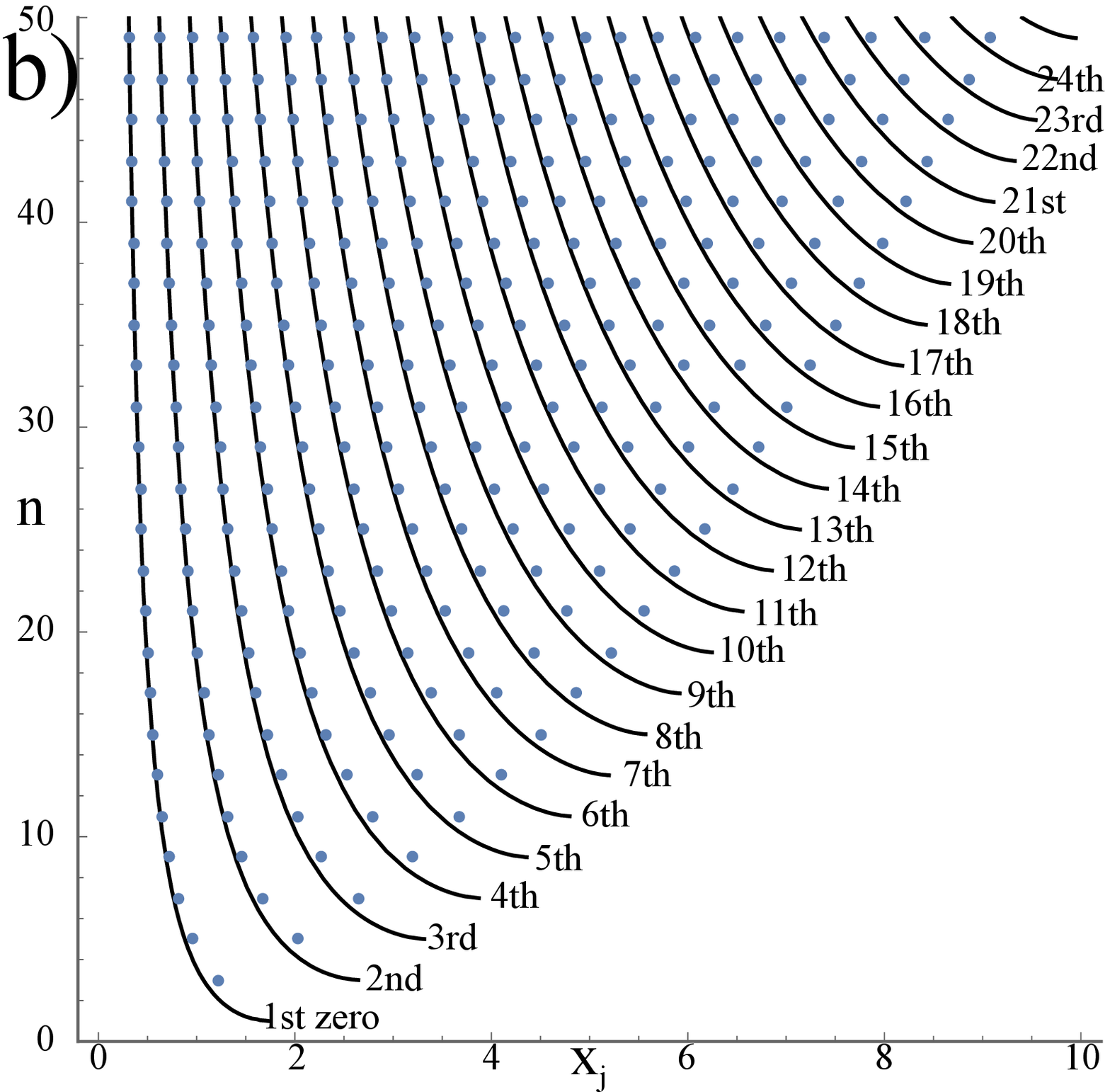}
\caption{Exact \(j\)th zeros of the Hermite polynomials \(H_n(x)\) for \(n\) a) even and b) odd compared to those obtained from solving eqs.~\ref{eq:evenzeros} and \ref{eq:oddzeros}.}
\label{fig:zeros}
\end{figure}

Solving these equations for \(x_j\) yields the approximate \(j\)th zero for the \(n\)th Hermite polynomial. The results for the zeros of the first \(50\) Hermite polynomials are compared to the exact zeros in fig.~\ref{fig:zeros}. In both cases, eqs.~\ref{eq:evenzeros} and \ref{eq:oddzeros} converge to the zeros of the Hermite functions quite quickly\footnote{J. Katriel, through correspondence, showed that eqns~\ref{eq:evenzeros} and \ref{eq:oddzeros} agree with the first asymptotic term from Dominici\cite{Dominici07} for \(n \rightarrow \infty\) for low \(j\) (not for maximal \(j\)). The latter result makes sense from the point of view that the maximal \(x_j\) is always close to the edge of the Hilbert space where the wavefunction goes to zero for any finite \(n\) whereas that of the harmonic oscillator decays forever. Eqs.~\ref{eq:evenzeros} and \ref{eq:oddzeros} do not agree with higher order terms (w.r.t. \(\frac{1}{n}\)) in Dominici's asymptotic expansion. }.

The finding that the boundaries of equal area partitions of growing circles correspond to the asymptotic zeros of the Hermite functions appears to be a novel one from a search of the literature. It is all the more surprising that the origin of this one-to-one correspondance stems from the Holstein-Primakoff representations for finite-dimensional spin systems. Furthermore, on a practical level, the apparently rapid convergence of these solutions suggests that they may be useful for more efficient determination of Hermite polynomial zeros for large-dimensional implementations of Gaussian quadrature. 

\section{Acknowledgments}

The author thanks Prof. J. Katriel for helpful comments on the manuscript.


\end{document}